\documentstyle[aps,twocolumn,prl,epsf]{revtex}

\newcommand{\barr}{\begin{array}}
\newcommand{\earr}{\end{array}}
\newcommand{\beq}{\begin{equation}}
\newcommand{\eeq}{\end{equation}}
\newcommand{\bea}{\begin{eqnarray}}
\newcommand{\eea}{\end{eqnarray}}
\newcommand{\beaa}{\begin{eqnarray*}}
\newcommand{\eeaa}{\end{eqnarray*}}

\makeatletter

\def\compoundrel#1\over#2{\mathpalette\compoundreL{{#1}\over{#2}}}
\def\compoundreL#1#2{\compoundREL#1#2}
\def\compoundREL#1#2\over#3{\mathrel
	{\vcenter{\hbox{$\m@th\buildrel{#1#2}\over{#1#3}$}}}}
\makeatother

\newcommand{\al}{\alpha}

\def\c+{c^{\dagger}}
\def\d+{d^{\dagger}}

\def\bra#1{{\langle #1 \vert}}
\def\ket#1{{\vert #1 \rangle}}

\begin{document}
\draft

\include{psfig}

\twocolumn[
\hsize\textwidth\columnwidth\hsize\csname @twocolumnfalse\endcsname

\title{Quantum Ferrimagnets}
\author{M.~Abolfath$^{\text a,b}$,  H.~Hamidian$^{\text c}$,
and A.~Langari$^{\text b,d}$}
\address{
$^{\text a}$Department of Physics, Indiana University, Bloomington, 
Indiana 47405}
\address{
$^{\text b}$Institute for Studies in Theoretical Physics and Mathematics
P.O. Box 19395-5531, Tehran, Iran}
\address{
$^{\text c}$Department of Physics, University of Illinois at Chicago, 
845 W. taylor Street, Chicago, IL 60607, USA}
\address{
$^{\text d}$Institute for Advanced Studies in Basic Sciences, 
Zanjan 45195-159, Iran}
\date{\today}
\maketitle

\begin{abstract}
\leftskip 2cm
\rightskip 2cm
We study quantum ferrimagnets in one, two, and three dimensions 
by using a variety of methods and approximations. These include:   
(i) a treatment based on the spin coherent state path-integral 
formulation of quantum ferrimagnets by taking into account the leading
order quantum and thermal fluctuations   
(ii) a field-theoretical (non-linear $\sigma$-model type)
formulation of the special case of one-dimensional quantum ferrimagnets 
at zero temperature (iii) an effective description in terms of dimers and  
quantum rotors, and (iv) a quantum renormalization group study of
ferrimagnetic Heisenberg chains. 
Some of the formalism discussed here can be used for a
unified treatment of both ferromagnets and antiferromagnets in the
semiclassical limit. 
We show that the low (high) energy effective Hamiltonian of a $(S_1, S_2)$ 
Heisenberg ferrimagnet is a ferromagnetic (antiferromagnetic) 
Heisenberg model. We also study the phase diagram of quantum ferrimagnets
in the presence of an external magnetic field $h$ ($h_{c1} < h < h_{c2}$ ) 
and show that the low- and the high-field phases correspond respectively 
to the classical N\'eel and the fully polarized ferromagnetic states. 
We also calculate the transition temperature for the 
Berezinskii-Kosterlitz-Thouless phase transition in the special case of 
two-dimensional quantum ferrimagnets. 
\end{abstract}


\ifpreprintsty\else\vskip1pc]\fi
\narrowtext

\section{Introduction}
\label{Intro}
Since the appearance of Haldane's conjecture \cite{haldane83} in 1983 the
study of quantum antiferromagnetic spin chains has been the subject of
much research activity in condensed matter physics. According to Haldane,
integer-spin Heisenberg antiferromagnetic chains
have a unique disordered ground state with a finite excitation gap,
while half-integer chains are gapless and critical.
The origin of the difference between half-integer and integer spin chains
can be
traced back to the topological ``$\theta$'' term in the effective
non-linear
$\sigma$-model
(NL$\sigma$M) description of antiferromagnetic spin chains and is believed
to be due to non-perturbative effects \cite{haldane83}.
Haldane's original predictions were based on large-spin (large-$S$)
arguments
and, although a general rigorous proof is still lacking, several
theoeretical
developments have helped to clarify the situation and there is now strong
experimental and numerical evidence in support of Haldane's claim  (see
Refs.~\cite{affleck89}, \cite{fradkin}, and \cite{auerbach} for details
and
further references).

Recently, antiferromagnetically coupled mixed-spin chains with an
alternating
array of two kinds of spins have also attracted interest among
researchers.
Integrable models of mixed-spin antiferromagnetic chains were constructed
by de~Vega and Woynarovich \cite{devega} and the
simplest case of such chains with spins $S=1$ and $1/2$ were subsequently
studied \cite{pati}. Since these integrable models are exactly solvable,
they are very useful for studying (quantum) statistical mechanical
properties. However, they involve complicated interactions and may be
inadequate for capturing the essential features of mixed-spin
chains with nearest-neighbor interactions. A similar situation arises in
the
study of Heisenberg antiferromagnets where Bethe ansatz integrable chains
\cite{kullish} with integer spin and no gap seem to contradict Haldane's
claim. These models include special generalized interactions and behave
quite differently from the pure Heisenberg chain.
Furthermore, from a more practical point of view, it is the mixed-spin
chains
with nearest-neighbor interactions which are supposed to describe
real ferrimagnetic compounds \cite{kahn}.

Several researchers have recently carried out analytical and numerical
studies
on the simplest quantum ferrimagnetic chains to obtain the low-temperature
properties of these systems \cite{pati}. As can be seen through
these works, although ferrimagnetic spin  chains exhibit
both ferromagnetic and antiferromagnetic features, they show some
peculiar,
and sometimes surprising,  features uncharacteristic of either the
ferromagnet or the antiferromagnet---an example being the
existence of gapless excitations with very small correlation length.
It is important to understand these features more clearly and much
theoretical, numerical, and experimental work lies ahead.

In this paper we present a systematic and detailed study of the effect of
both thermal and quantum fluctuations in one-, two-, and three-dimensional
quantum
ferrimagnets with nearest neighbor interactions and
with arbitrary (alternate) spins $S_1$ and $S_2$
on the two sublattices of a bi-partite lattice.
The outline of our work is as follows:
in Sec.~II we present a formulation based on 
the spin coherent state representation and discuss the role of
fluctuations
in quantum ferrimagnets.
Using the formulation described in Sec.~II, we continue in Sec.~III with
explicit analytical calculations of magnetization in quantum ferrimagnets
at finite temperature and discuss the low- and
high-temperature limits. In Sec.~IV we re-formulate the quantum
ferrimagnetic spin chain as a $1+1$ dimensional quantum field theory
and explore its novel features and
(dis)similarities
(such as the existence of a topological $\theta$ term)
with antiferromagnetic Heisenberg spin chains.
In Sec.~V the physical properties of quantum ferrimagnets, as described 
by the continuum NL$\sigma$M formulation, are examined further  
and the quantum phases of dimerized states are discussed in some detail. 
The effective Hamiltonian based on a description in terms of dimers,    
and a detailed quantum renormalization group study of ferrimagnetic 
chains is presented in Sec.~VI.
We will explicitly show that in the low-energy limit quantum ferrimagnets
can be effectively descibed as ferromagnets and a cross-over to an 
antiferromagnetic phase takes place as the temperature is increased.  
We conclude in Sec.~VII with a summary of
our work and closing remarks and suggestions for future directions.

\section{Spin-Wave Theory and the Semiclassical Limit }
\label{Class}
In this section we present a formulation based on the semiclassical
expansion of the spin path intergral to obtain the phase
diagram of quantum ferrimagnets with nearest-neighbor interactions in
$d \leq 3$ dimensions.

To obtain the Hamiltonian for the ferrimagnet we begin by dividing the
spin system into two (arbitrary) sublattices $A$ and $B$ with spins $S$
and $KS$ on each sublattice respectively, where $|K| \equiv S_2/S_1$ is a
number that specifies the relative magnitude of the spins
$S_1(=S)$ and $S_2(=KS)$ on the $A$ and $B$ sublattices.
Note that the sign of $K$ specifies the relative direction of the
spins on sublattices $A$ and $B$.
The classical Hamiltonian can be written as
\beq
H_{\rm classical} = K J S^2 \sum_{<i,j>} {\bf m}_i \cdot {\bf m}_j,
\label{R1}
\eeq
where $J > 0$. We have used the spin coherent state representation to write the
spin operators on the $A$ and $B$ sublattices as
${\bf S}_i = S {\bf m}_i$ and ${\bf S}_i = KS {\bf m}_i$ respectively,
with ${\bf m}_i$ a 3-component classical unit vector \cite{fradkin}.
As can be easily seen, the Hamiltonian (\ref{R1}) exhibits a duality
symmetry between the large and small $|K|$ limits.
One may increase the magnitude of spins on sublattice $B$,
or equivalently reduce the magnitude of spins on sublattice $A$.
In general the Hamiltonian (\ref{R1}) is invariant under the
transformation $K \rightarrow 1/K$ and $S \rightarrow KS$.
There is thus a duality symmetry between the large and small $|K|$
values and, in this sense, the ferrimagnetic spin system is self-dual:
The properties of the system at large
$|K|$ can be mapped onto those of the dual system at small $|K|$.
This duality symmetry is similar to that of the
two-dimensional (2D) Ising
model, as first discussed by Kramers and Wannier \cite{kramers}, where a
one-to-one correspondence between the high and low temperature limits of
the partition function can be established.
In the ferrimagnetic case, as described by (\ref{R1}), one may think
of $K$ as (a fictitious) temperature
and make an analogy with the familiar 2D Ising model.
A consequence of this self-duality is that the ferrimagnet must exhibit
identical behavior in the $|K|=\infty$ and $K=0$ limits.

On general grounds, the sign of $K$ specifies the long-range (classical)
behavior of the system, i.e. the ground state has either a non-zero
staggered, or a fully polarized magnetization for positive or negative
$K$'s respectively.
When $K > 0$, fluctuations around the classical solution can be taken into 
account by reparameterizing the unit vectors ${\bf m}_i$ as \cite{SMGNotes}
\begin{mathletters}
\label{R2}
\beq
{\bf m}_i = \sqrt{1 - \overline{\psi}_i \psi_i} \hat{z} + \psi_{xi}
\hat{x}
+ \psi_{yi} \hat{y} \hspace{12mm} i \in A,
\label{R2a}
\eeq
\beq
{\bf m}_i = -\sqrt{1 - \overline{\psi}_i \psi_i} \hat{z} + \psi_{xi}
\hat{x}
+ \psi_{yi} \hat{y} \hspace{1cm} i \in B,
\label{R2b}
\eeq
\end{mathletters}
\noindent
where $\psi_i = \psi_{xi} + i \psi_{yi}$ is a complex scalar field. The
macroscopic spontaneous magnetization, $M^K_z$, along
an arbitrary axis, e.g. $\hat{z}$, can be taken as the order parameter for
the ferrimagnet and is given by
\beq
M^K_z \equiv  \langle S_{zi} \rangle_{i \in A} - {\rm sgn}(K)
\langle S_{zi} \rangle_{i \in B}.
\label{R3}
\eeq
The lowest-lying collective modes of the system can be taken into account
by
linearizing the Hamiltonian in terms of the fluctuation fields, $\psi_i$
\bea
H_{\rm eff} &\equiv& H_{\rm classical} - E_0  \nonumber\\&& =
\frac{KJS^2}{2} \sum_{i\in A} \sum_{j\in B}
(\overline{\psi}_i \psi_i +\overline{\psi}_i \psi_j
+\overline{\psi}_j \psi_i +\overline{\psi}_j \psi_j),
\label{R4}
\eea
where $E_0$ is the ground state energy of the system. The dynamical part
of the action consists of the quadratic (Wess-Zumino) term \cite{fradkin}
\beq
S_{WZ} = \frac{\hbar S}{2} \int_0^{\hbar\beta} d\tau \left\{\sum_{i\in A}
\overline{\psi}_i \partial_\tau \psi_i - K \sum_{i\in B}
\overline{\psi}_i \partial_\tau \psi_i\right\},
\label{R5}
\eeq
which reduces to the corresponding Wess-Zumino term for the familiar
quantum (anti)ferromagnet when $K=\pm 1$. 
It is important to note that
when $K < 0$, one must use Eq.~(\ref{R2a}) for the unit vector ${\bf m}_i$ on
the sublattice B. This results in a minus sign in front of the intra-site
interaction terms in the Hamiltonian (\ref{R4}).

The contribution of the low-lying fluctuations to the effective action can
be calculated by Fourier expanding the field $\psi$ which yields
\beq
S_{\rm eff} = \frac{S\hbar\beta}{2} \sum_k \sum_n
\left(\begin{array}{c}
\overline{\alpha}_n(k) \;\; \overline{\beta}_n(k)\end{array}\right)
{\cal S}_n(k)
\left(\begin{array}{c}
\alpha_n(k) \\ \beta_n(k)\end{array}\right),
\label{R6}
\eeq
with
\beq
{\cal S}_n(k) =
\left(\begin{array}{cc}
-i\hbar\omega_n + KJSz & KJS\overline{\gamma}(k) \\
KJS\gamma(k) & iK\hbar\omega_n + KJSz
\end{array}\right).
\label{R7}
\eeq
Here $z$ is the coordination number,
$\gamma(k) = \sum_\xi e^{-i\vec{\xi} \cdot {\bf k}}$, and $\vec{\xi}$
is a vector that connects the nearest-neighbor spins, and
$\omega_n = 2\pi n/\hbar\beta$
denotes the bosonic Matsubara frequencies.
The dispersion relation for the collective modes is then obtained by
evaluating the poles of ${\cal S}_n(k)$:
\bea
\hbar\Omega^{\pm}(k) &=& \frac{K-1}{2} JSz \nonumber\\&& \pm
\frac{1}{2}JSz
\sqrt{(K - 1)^2 + 4 K[1-\frac{1}{z^2} |\gamma(k)|^2]}.
\label{R8}
\eea
and then by applying the analytical continuation of the Matsubara
frequencies $i\hbar\omega^{\pm}_n \rightarrow \pm\hbar\omega^{\pm}_k$.
Note that Eq.(\ref{R8}) gives the dispersion relations for $K > 0$. 
For $K < 0$, it is
easy to find the dispersion relation of the spin waves, by exploiting
$i\hbar\omega^{\pm}_n \rightarrow \hbar\omega^{\pm}_k$.

This equation explicitly exhibits the self-duality of the quantum
ferrimagnet
that was mentioned earlier. As can be readily seen from Eq.~(\ref{R8}),
the $K \rightarrow 1/K$ and
$S \rightarrow KS$ limits yield similar
dispersion relations, thus verifying the aforementioned self-duality of
the
quantum ferrimagnet in $K$-space.
In general, ${\cal S}_n(k)$ has two positive and negative poles
depending on the sign of $K$.
For $K=1$ we find the classical N\'eel state with gapless spin
waves, $\Omega_k^{\pm} \propto \pm k$. In this case both negative
and positive poles are present. In Sec.~\ref{ZPQF} we will show that
the negative poles contribute to the zero point quantum fluctuations
(ZPQF's)
at zero temperature and result in a staggered magnetization, in agreement
with calculations using the Holstein-Primakoff transformation.
For $K=-1$, in which case all spins are aligned along the same direction,
the
poles are positive and there are no zero point quantum fluctuations.
The poles in this case consist of gapless ($\Omega_k^+ \sim k^2$)
and gapped ($\Omega_k^- \sim \Delta - k^2$) modes since the lattice contains
a double-basis Brillouin zone.
Treating $K$ as a continuous parameter
(see also comments in the paragraph following
Eq.~(\ref{ferrimagnetaction}))
for tuning phase transitions, we see that
at $K=0$ (and zero temperature) the ground state of the system changes
abruptly and the negative poles change sign. Quantum
fluctuations are responsible for this change and
a continuous phase transition takes place at $K=0$.
In the latter, the system is completely disjointed, and the spins
are fully uncorrelated. In this case the system is in the disordered 
phase due to the ZPQF's.
For $K > 0$ there are two poles: A negative pole, which behaves like $k^2$
as
$k$ tends to zero, and a positive pole with a gap. For small wave
vectors the positive pole approaches the gap 
$\Delta (= JSz|1-K|)$ 
as $k^2$ approaches zero.
In this case the ZPQF's still persist. Since phase transitions at non-zero
temperature $T$ depend on 
both $T$ and $\Delta$, 
for $T \ll \Delta$ the gapped mode can be truncated from the
Hilbert space and the long-wavelength behavior of the system can be
effectively obtained from the $k^2$-gapless mode. In this case
the ferrimagnet mimics the ferromagnet in the long distance limit,
but in the presence of ZPQF's.
In the other extreme limit when $T \gg \Delta$, the gap is
negligible compared to thermal fluctuations and the system behaves like
an antiferromagnet. We therefore expect a cross-over from the
pseudo-ferromagnetic state to the antiferromagnetic state
at $T_c (\propto |1-K|)$. To see the physical reason for the
presence of the gap more clearly, let us transform Eq.~(\ref{R8}) to
a linearized (space-time) equation of motion
\bea
\left\{-\partial^2_\tau - JSz(1-K) \partial_\tau + 2K(JSz)^2 a^2_0
\nabla^2
\right\} \psi = 0,
\label{R8a}
\eea
where $a_0$ is the lattice spacing and $\tau$ is the imaginary time.
As we can see, the second term in
this equation is linear in the time derivative and therefore implies a
time-reversal symmetry breaking. This term is absent when $K=1$ 
(or when $T \gg \Delta$)
which corresponds to the antiferromagnet. The significance of
this
time-reversal symmetry breaking term lies in the fact that it can be
effectively considered as arising from a local magnetic field that
changes the magnitude
of the spins on the $B$ sublattice. As a result, according to this
picture,
the gap can be interpreted as the Zeeman energy cost for a spin flip.

In the following section we explicitly
calculate the magnetization for quantum ferrimagnets to leading order by
taking
the effect of fluctuations into account. These calculations further
support the
above
ideas.
\section{Magnetization}
\label{ZPQF}
The effect of fluctuations on magnetization can be computed by using
the (finite temperature) Green's functions
\beq
G_i^{\pm}(\tau) \equiv 
\lim_{\tau\rightarrow 0^\mp} \langle T_\tau
\psi_i(\tau) \overline{\psi}_i(0) \rangle, \hspace{10mm} {\rm if}~~ K > 0
\eeq
and 
\beq
G_i^{\pm}(\tau) \equiv 
\lim_{\tau\rightarrow 0^-} \langle T_\tau
\psi_i(\tau) \overline{\psi}_i(0) \rangle, \hspace{10mm} {\rm if}~~  K < 0 
\label{R10}
\eeq
where $+(-)$ labels the sublattice A(B) to which the $i$th site belongs,
the
ensemble average is
denoted by $\langle \cdots \rangle$, and $T_\tau$ is the time-ordering
operator. The magnetization can be found from (\ref{R3}) with
$\langle S_{zi} \rangle$
given by
\beq
\langle S_{zi} \rangle = \left\{\begin{array}{c}
S - \frac{S}{2} \langle \overline{\psi}_i \psi_i \rangle \hspace{19mm}
i\in A
\\    \\
-KS + \frac{KS}{2} \langle \overline{\psi}_i \psi_i \rangle \hspace{1cm}
i\in B \end{array}\right.
\label{R9}
\eeq
By applying the standard techniques, we can use Eqs.~(\ref{R7}) and
(\ref{R8})
to evaluate $G^\pm_i$ by integrating out fluctuations
and summing over the Matsubara frequencies,
\begin{mathletters}
\label{R11}
\bea
G^+_i = \frac{2}{KNS}&& \sum_k \left\{
\frac{K\hbar\Omega^+_k + KJSz}{\hbar\Omega^+_k - \hbar\Omega^-_k}
n_B(\beta\hbar\Omega^+_k)  \right.\nonumber\\&& \left. +
\frac{K\hbar\Omega^-_k + KJSz}{\hbar\Omega^-_k - \hbar\Omega^+_k}
n_B(\beta\hbar\Omega^-_k)\right\},
\label{R11a}
\eea
\bea
G^-_i = \frac{2}{KNS}&& \sum_k \left\{
\frac{-\hbar\Omega^+_k + KJSz}{\hbar\Omega^+_k - \hbar\Omega^-_k}
[n_B(\beta\hbar\Omega^+_k) + 1]  \right.\nonumber\\&& \left. +
\frac{-\hbar\Omega^-_k + KJSz}{\hbar\Omega^-_k - \hbar\Omega^+_k}
[n_B(\beta\hbar\Omega^-_k) + 1]\right\}.
\label{R11b}
\eea
\end{mathletters}
It is important to note that $G^+_i$ is the Green's function of the
sublattice A for any positive or negative $K$'s,
but $G^-_i$ is the Green's function of the sublattice B
for $K > 0$. One may easily show that the Green's function
of the sublattice B for $K < 0$ can be obtained by Eq.(\ref{R11b}) if
$n_B(\beta\hbar\Omega_k^\pm) + 1 \rightarrow n_B(\beta\hbar\Omega_k^\pm)$.
It is easy to show that the magnetization (along the $z$-axis)
per unit cell at $T=0$ satisfies the 
sum rule: $M_0 \equiv \langle S_{zi} \rangle_{i\in A} 
+ \langle S_{zi} \rangle_{i\in B} = |S_1 - S_2|$ when $K > 0$.
In this case, both the magnetization and the staggered magnetization
are non-vanishing simultaneously.
The non-zero $M_0$ changes the low energy physics of the ferrimagnets
abruptly, i.e., they behave like the ferromagnets rather than
antiferromagnets. we will come to this point in Sec. \ref{NLSM}. We
therefore are led to considering $M_0$ as an additional order parameter
for the ferrimagnets, i.e., a ferrimagnet with $K > 0$ is fully described
by two non-zero order parameters $M^{Neel}_z$ and $M_0$.
Note that for $K < 0$ there is no difference between our definition
for the order parameter $M_z^K$ and the usual magnetization $M_0$.

Before proceeding further with the calculation of magnetization for the
ferrimagnet, and in order to check the consistency of the formalism
adopted
here, let us first examine some well-known results.
For $K=-1$ we have a ferromagnet and
the magnetization can be obtained from
Eqs.~(\ref{R11}), (\ref{R9}), and (\ref{R3}) which give
\bea
M_z = 2S - \frac{1}{N} \sum_k \left\{ n_B(\beta\hbar\Omega^+_k) +
n_B(\beta\hbar\Omega^-_k)\right\},
\label{R12}
\eea
The asymptotic limit of Eq.~(\ref{R12}) at low temperature
($\Omega^+_k \ll T \ll JSz$)
can be studied by introducing the infrared cutoff $\Lambda^{-1}$ (the
typical system size is characterized by $\Lambda$). The second term in
Eq.~(\ref{R12}) is irrelevant in the long-wavelength limit and we obtain:
\beq
M_z - 2S \propto \left\{\begin{array}{cccc}
-\frac{T}{JSz} \Lambda & \; & \; & d=1 \\  \\
-\frac{T}{JSz} \log \Lambda & \; & \; & d=2 \\ \\
-0.02 ~ \zeta\left(\frac{d}{2}\right)
\left(\frac{T}{JSz}\right)^{d/2} & \; & \; & d=3
\end{array}\right.
\label{R15}
\eeq
where $\zeta(x)$ is the Riemann zeta function.
We can directly see that there are no ZPQF's at $T=0$ in any dimension and
that for any low $T$ at $d=1, 2$ the correction to magnetization diverges
due to severe thermal fluctuations and there can be no long-range order.
This is in accord with the Mermin-Wagner theorem \cite{Mermin-Wagner},
which
is applicable to one- and two-dimensional systems with a continuous
symmetry.
However, in $d=3$ the correction due to fluctuations is finite and there
{\em can} be long-range order, with the order-disorder phase transition
taking place at $T_c \sim zJS^{5/3}$.

The other familiar case corresponds to $K=1$ where the classical ground
state
is the N\'eel state with
$\hbar\Omega_k^\pm = \pm JSz \sqrt{1 - |\gamma_k|^2/z^2}$.
The staggered magnetization in this case can be found from
Eqs.~(\ref{R11}), (\ref{R9}), and (\ref{R3})
\bea
M^{Neel}_z &=& 2S + 1 - \frac{1}{N} \sum_k
\frac{1}{\sqrt{1 - \frac{1}{z^2}|\gamma(k)|^2}} \nonumber\\&& \times
\coth\left(\frac{\beta JSz}{2} \sqrt{1 -
\frac{1}{z^2}|\gamma(k)|^2}\right),
\label{R16}
\eea
and the correction due to fluctuations at low temperatures
is given by:
\beq
M^{Neel}_z - 2S - 1 \propto \left\{\begin{array}{cccc}
-\alpha_1 - \frac{T}{JSz} \Lambda & \; & \; & d=1 \\  \\
-\alpha_2 - \frac{T}{JSz} \log \Lambda & \; & \; & d=2 \\ \\
-\alpha_3 - \frac{0.006}{d} \left(\frac{T}{JSz}\right)^2 & \; & \; & d=3
\end{array}\right.
\label{R17}
\eeq
where $\alpha_d = N^{-1} \sum_k (1 - |\gamma(k)|^2/z^2)^{-1/2}$ in $d$
dimensions. As we can see, that the ZPQF's are present at $T=0$ and, 
similarly to a ferromagnet, long range order is destroyed at
$d=1, 2$ at any
temperature; a result which is also expected from the Mermin-Wagner
theorem.

The results obtained above for the two cases with $K=1$ and $K=-1$ agree
with
those that have been established previously by using the standard
spin-wave
approximation
\cite{auerbach}.
(We should like to
note that although the behavior of (at least) one-dimensional (1D)
quantum antiferromagnets at zero $T$
requires a treatment that goes beyond the spin-wave approximation, the
finite
$T$ results discussed above are valid independently of these
modifications.)

We shall now treat $K$ as an adjustable parameter to study the general
case
of the quantum ferrimagnet. The fluctuation-corrected (staggered)
magnetization can be found as before and one obtains:
\bea
M^K_z &=& (1 + |K|)S + \theta(K) - \frac{1}{N}  \sum_k g(K, k)
\nonumber\\&& \times
\left\{n_B(\beta\hbar\Omega^+_k)
- {\rm sgn}(K) n_B(\beta\hbar\Omega^-_k)\right\},
\label{R18}
\eea
where $\theta(K)$ is the Heviside function and 
\bea
g(K, k) =
\left\{\begin{array}{cccc}
1 & \; & \; & K < 0 \\  \\
\frac{1+K}{\sqrt{(1-K)^2+4K(1-|\gamma(k)/z|^2)}} & \; & \; & K > 0
\end{array}\right.
\label{R18a}
\eea
It is easy to check that Eq.~(\ref{R18}) reduces to Eqs.~(\ref{R15}) and
(\ref{R17}) when $K=-1,1$ respectively.
To simplify, let us rewrite
Eq.~(\ref{R18}) as
\beq
M^K_z = (1+|K|)S + \theta(K) - \alpha_d - \Delta m^d_z,
\label{R19}
\eeq
where the identity $n_B(-|x|) = -1 - n_B(|x|)$ has been used. Here
\beq
\alpha_d =
\left\{\begin{array}{cccc}
0 & \; & \; & K < 0 \\  \\
\frac{1}{N} \sum_k g(K, k) & \; & \; & K > 0
\end{array}
\right.
\label{R20}
\eeq
which contributes to the ZPQF's at $T=0$, and
\bea
\Delta m^d_z =
\frac{1}{N} \sum_k g(K, k) [n_B(\beta\hbar\Omega^+_k)
+ n_B(\beta\hbar|\Omega^-_k|)].
\label{R21}
\eea
Note that for $K<0$ both $\Omega^\pm_k$ are positive and $\alpha_d =0$.

The temperature dependence of the ferrimagnetic order parameter $M^K_z$
resides in $\Delta m^d_z$.
We can therefore learn about the behavior of the system by studying
Eq.~(\ref{R21}) for general $T$ and $K$.
Similarly, the temperature dependence of the (usual) 
magnetization $M_0$ (for ferrimagnets with $K > 0$) is:
\beq
M_0 = |S_1 - S_2| - \frac{1}{N} \sum_k \left| n_B(\beta\hbar\Omega^+_k)
- n_B(\beta\hbar|\Omega^-_k|)\right|.
\label{mm1}
\eeq
As can be seen from this equation, there is no ZPQF contribution to the
$M_z$ at $T=0$, and 
that for any low $T$ at $d=1, 2$ the correction to magnetization diverges
due to severe thermal fluctuations and there can be no long-range order
(the Mermin-Wagner theorem \cite{Mermin-Wagner}), similar to (\ref{R15}).

As has already been mentioned
in Sec.~\ref{Class}, for $K > 0$
the compitition between the temperature and the energy gap in the system
results in different long-range behavior and,
since $\Delta \propto |1 - K|$, we expect the compitition between $T$
and $|1 - K|$ to determine the order-disorder phase transitions in the quantum
ferrimagnet. For instance, the high temperature limit is
equivalent to the small $|1 - K|$ (i.e. $K \sim 1$) limit.
Let us therefore investigate the asymptotic behavior
of the staggered magnetization in both the low and high temperature
limits,
for $K > 0$.
>From now on until the end of the paper we concentrate our attention
to the ferrimagnets with $K > 0$.

{\em The low temperature limit}. This limit occurs in the range
$|\Omega^-_k| \ll T \ll |1 - K|$ and
one may neglect the $1 - |\gamma_k|^2/z^2$ dependence in Eqs.~(\ref{R8})
and (\ref{R20}) compared to $(1 - K)^2$ in the
$k \rightarrow 0$ limit. As a result, $\alpha_d \simeq 2 (1+K)/|1-K|$ and
one
obtains a residual contribution to ZPQF's with no dependence on the
dimensionality of the system.
In this limit the action contains two poles, $\Omega^+_k =
JSz(|1-K| + 2K k^2/|1-K|)$ and
$|\Omega^-_k| = 2KJSz k^2/|1-K|$ as $k \rightarrow 0$, and we find:
\beq
\Delta m^d_z
\propto \left\{\begin{array}{ccc}
\frac{|1-K|}{2K}\frac{T}{JSz} \Lambda & \; & d=1 \\  \\
\frac{|1-K|}{2K}\frac{T}{JSz} \log \Lambda & \; & d=2
\end{array}\right.
\label{R22}
\eeq
and
\beq
\Delta m^d_z =
0.02 \zeta\left(\frac{3}{2}\right) (1+K)|1-K|^{3/2}
\left(\frac{T}{2KJSz}\right)^{3/2},
\label{R23}
\eeq
when $d=3$.
Note that the contribution of the gapless modes ($\Omega^-_k$) is
responsible for the infrared
divergences in (\ref{R22}) and softening of the long-range order of the
system in low-dimensions.
The behavior of the system is different from an antiferromagnet
and the temperature dependence of its staggered magnetization is the same
as
that of the ferromagnet (see Eq.~(\ref{R15})).
This corresponds to the pseudo-ferromagnet phase in Fig.~\ref{Fig1}.

{\em The high temperature limit}. This limit occurs in the range
$T \gg |1 - K|$ or $K \sim 1$ and,
in contrast to the previous case, here we may neglect $(1 - K)^2$ in
Eqs.~(\ref{R8}) and (\ref{R20}) as compared to $1 - |\gamma_k|^2/z^2$.
[In Sec.~IV we will explicitly show that at $d=1$ the behavior of quantum 
ferrimagnets in the high temperature (high energy) limit is 
governed by antiferromagnetic exchange interactions. 
Although the proof of this statement for dimensions $d>1$ is still lacking,
we believe the qualitative picture is the same at higher dimensions, where 
a mean-field description becomes more reliable.]
As expected, in this limit the system behaves more like an antiferromagnet
with the staggered magnetization given by Eq.~(\ref{R17}).
We label this state as the antiferromagnetic phase in Fig.~\ref{Fig1}.
To understand the phase boundary in Fig~\ref{Fig1} better, let us recall
Eq.~(\ref{mm1}). As can be seen from this equation, at zero temperature 
$M_0=0$ when $K=1$, but as the temperature is increased 
$M_0$ vanishes at $K=K_c$ where $K_c \neq 1$ and $M_0$ is suppressed by
the thermal fluctuations. Since in this region
(where $M_0 = 0$),
the (staggered) antiferromagnetic order parameter, $M^{Neel}$, 
is non-vanishing, the system is in an antiferromagnetic phase. 
From Eq.~(\ref{mm1}) one finds that $T \propto |1 - K_c| S$ is the temperature
at which a cross-over from the pseudo-ferromagnetic phase into the 
antiferromagnetic phase takes place.  

The order-disorder phase boundaries can now be identified in terms of the
staggered magnetization $M^K_z$.
For $d=3$, and far enough from $K=1$, the critical temperature is given by
\beq
T_c \propto JSz |K|/|1-K|^{1/3}.
\label{R24}
\eeq
Note that close to the $K=1$ limit the critical temperature can be
obtained
from Eq.~(\ref{R17}) by solving $M^{Neel}_z = 0$. Fig.~\ref{Fig1}
shows
the qualitative phase diagram for the quantum ferrimagnets. The phases
denoted
by antiferromagnet and pseudo-ferromagnet are separated by 
$T_c \sim |1-K|$ as discussed above. Due to the Mermin-Wagner theorem,
and as has been explicitly calculated here, there can be no long-range
order
at non-zero temperatures when  $d=1, 2$ and hence no phase transitions
either.
It is noteworthy to emphesize that at $d=2$ the staggered magnetization
and the usual magnetization are 
diverged logarithmically, as one can see from Eq.(\ref{R22}). This is the
reminiscent of the low-temperature quasi-long range order phases, where
the vortices bind in pairs. We therefore expect a non-zero temperature
Berezinskii-Kosterlitz-Thouless phase transition \cite{KT} takes place
when
the quasi-long range order of the 2-dimensional ferrimagnets is broken
(when the vortices are deconfined). We will come to this point in
Sec. \ref{QPT}.
At $K=1$ the ferrimagnetic gap ($\Delta \propto |1-K|$) collapses and the
N\'eel state appears.
The phases of the system now depend on $S$, i.e. for half-integer $S$
the model is critical and the
correlation function of the staggered magnetization has a power law decay,
while for integer $S$ the spin chain is disordered with a Haldane gap.
This completes our study of quantum ferrimagnets at zero- and non-zero
temperature within the semiclassical framework of spin-wave 
approximation. Our calculations include the effects of 
both quantum and thermal fluctuations to leading order. 

\section{The Continuum limit}
\label{NLSM}
In this section we re-formulate the $K > 0$ and the zero-temperature
1D quantum ferrimagnetic chain defined by
(\ref{R1}) as a quantum field theory in $1+1$ dimensions.
As will be seen below,
the continuum action obtained here is a generalization of the familiar
(NL$\sigma$M) formulation
of the 1D Heisenberg antiferromagnet.
There are many advantages in using
continuum field theoretical descriptions of (quantum) spin systems. These
include the
possibility of studying non-perturbative aspects of the model and making
the existence (or lack of) certain symmetries manifest. A notable example
is,
of course, the original mapping of the 1D Heisenberg antiferromagnet
onto the NL$\sigma$M which led Haldane to predict that
integer-spin chains have a finite spin excitation gap in their spectrum
and
half-integer spin chains do not. This difference between integer and
half-integer spin chains is understood to
be due to non-perturbative (topological) effects and could not have been
obtained by using the standard spin-wave theory alone.

As before, we use the spin coherent state representation to study the
ferrimagnetic chain in the continuum limit.
Assuming a total number of $2N$ sites and periodic boundary
conditions, it is straightforward to obtain the action
\bea
{\cal S}[{\bf m}] =&& S \sum_{i\in A} S_{WZ}[{\bf m}_i] - KS \sum_{i\in B}
S_{WZ}[{\bf m}_i]\nonumber\\&&
+\int_{0}^{\hbar\beta} d\tau ~KJS^2 \sum_{<i,j>} {\bf m}_i \cdot
{\bf m}_j\nonumber\\
  =&& iS \Upsilon_A - iKS \Upsilon_B + {\cal H}_{\rm classical},
\label{action1}
\eea
where $\Upsilon_{A(B)}$ is the (topological) Berry phase associated with
the N\'{e}el field on the sublattice $A(B)$ and
\beq
S_{WZ} \equiv \int_{0}^{1} dt \int_{0}^{\hbar\beta} d\tau
{\bf m}(t,\tau)\cdot[\partial_\tau{\bf m}(t,\tau) \times \partial_t
{\bf m}(t,\tau)],
\label{swz}
\eeq
is the Wess-Zumino term. The action ${\cal S}$ is invariant under
the transformations $K \rightarrow 1/K$, $S \rightarrow KS$ and the
interchanging the sublattices $A$ and $B$.
(Note that the Wess-Zumino term has already been introduced in
(\ref{R5}), albeit in a different form. The formalism used in Sec.~I,
in which complex scalar fields $\psi_i$ were used, is akin to the
Schwinger
boson representation and is very
useful at both zero and finite temperature. However, for our present
($T=0$) purposes it is most convenient to use the formalism adopted here.)
The significance of the Wess-Zumino term in the continuum treatment
of the Heisenberg antiferromagnet, i.e. when $K=1$, can be
understood as follows \cite{fradkin}: On a
bipartite lattice, and starting with a (classical) N\'{e}el state around
which the fluctuations are to be taken into account, the staggered
configuration
leads to a change of sign of the exchange term in the action
(\ref{action1}) to
that of a ferromagnet. It is thus the staggered sum of the Wess-Zumino
terms, a purely quantum mechanical contribution, which distinguishes
ferromagnets from antiferromagnets.
In order to take the effect of fluctuations into account one writes
${\bf m}_i$ as the sum
of a slowly varying part, the {\em N\'{e}el} field ${\bf n}_i$, which
represents the order parameter, and a
small rapidly varying part, the transverse {\em canting} field ${\bf
l}_i$,
which roughly represents the average spin
\cite{affleck88}
\beq
{\bf m}_i = {\bf n}_i + (-1)^i ~ a_0 ~ {\bf l}_i,
\label{spinfield}
\eeq
with the constraints ${\bf n}_i^2=1$ and ${\bf n}_i\cdot{\bf l}_i=0$.
In the large-$S$ limit one obtains the
well-known Haldane's mapping \cite{haldane83} of the Heisenberg
antiferromagnet onto the (effective) $O(3)$ non-linear $\sigma$-model
by integrating out the canting fields (the fast modes). The resulting
action
consists of a NL$\sigma$M term through the (Euclidean) Lagrangian density
\beq
{\cal L}_{\rm NL\sigma M}({\bf n}) = \frac{1}{2 g}
\left[\frac{1}{v_s} (\partial_\tau {\bf n})^2 + v_s
(\partial_x {\bf n})^2 \right],
\label{nlsm1}
\eeq
where $g=2/S$, $v_s=2a_0JS$, as well as a topological surface term
expressed in terms of the two-dimensional Pontryagin density of the
hedgehog fields
\beq
{\cal L}_{\rm top}({\bf n}) =
i \frac{\theta}{8 \pi} \epsilon_{i j} {\bf n} \cdot
(\partial_i {\bf n}
\times \partial_j {\bf n}),
\label{top1}
\eeq
where $\theta=2\pi S$, and $\epsilon_{i j}$ is
the two-dimensional antisymmetric Levi-Civita tensor.
This term results in a purely topological contribution
to the action through $\theta W$, where $W$ (the winding number) is an
integer which appears through $e^{i\theta W}$ in the path integral.
(Note that the topological term (\ref{top1}) is absent when $d > 1$
\cite{Haldane88}.)
Since $\theta$ appears in the path integral through $e^{i \theta W}$, it
matters only modulo $2 \pi$, and therefore $\theta = 0, \pi$ for integer
and half-integer spin chains respectively.
Haldane conjectured that all half-integer (integer) spin chains are
gapless (gapped) \cite{haldane83}.
It has been known for sometime \cite{hamer} that the NL$\sigma$M with
$\theta=0$ is always
disordered and has exponentially decaying correlations in the
long-wavelength
(strong coupling) limit. At $\theta = \pi$, since the spin-1/2 chain is
known to
be gapless from Bethe's exact solution, one expects
algebraically decaying correlations, provided that the
(large-$S$) mapping onto the sigma model remains valid down to $S=1/2$.

We will now examine the more general model with $K \neq 1$ which
corresponds
to the ferrimagnetic chain.
In this case the action (\ref{action1}) can be rewritten as
\bea
{\cal S}[{\bf n}] &=&
\frac{1-K}{2} iS (\Upsilon_A + \Upsilon_B)
\nonumber \\&& +
\frac{1+K}{2} iS (\Upsilon_A - \Upsilon_B)
+ {\cal H}_{\rm classical},
\label{action2}
\eea
and in the continuum limit we obtain the Euclidean action:
\begin{eqnarray}
{\cal S}[{\bf n}] = \int_0^{\hbar\beta}d\tau \int dx {\cal L}({\bf n}),
\label{action3.2}
\eea
where
\begin{eqnarray}
{\cal L}({\bf n}) = i M_0 {\bf A}({\bf n}) \cdot \partial_\tau {\bf n}
+ {\cal L}_{\rm NL\sigma M}({\bf n}) + {\cal L}_{\rm top}({\bf n}).
\label{ferrimagnetaction}
\eea
The
first term in Eq.~(\ref{ferrimagnetaction}) is the usual (dynamical)
Berry's
phase of a quantum ferromagnet, with $M_0 \equiv |1-K|S/a_0$ the
magnetization per unit cell (pair of sites).
It is this term that results in the ferromagnetic branch of
the spin waves and corresponds to the trajectory of spin over a closed
orbit on the unit sphere in the presence of a unit magnetic monopole at
the center.
The contribution of the first term in ${\cal L}({\bf n})$ is equivalent to
the area enclosed by this trajectory and since either of the smaller or
the
larger enclosed areas on the unit sphere must lead to the same Berry's
phase,
the magnetic moment per unit cell,
i.e. $2 M_0 a_0$, must be quantized with integral values \cite{read}.
A corollary which follows from this last observation is that the
parameter $K$ is also quantized.
It is interesting to note that although at the spin-wave level
one can consider the parameter $K$ as a continuous parameter,
when including
topologically nontrivial spin histories, the theory is
inconsistent unless $K$ is quantized.
[Note, however, that in the large-$S$ limit $K$ can be considered as {\em
nearly
continuous} as long as $S$ is large on both
sublattices.]
The remaining terms in Eq.~(\ref{ferrimagnetaction}) are the NL$\sigma$M
\bea
{\cal L}_{\rm NL\sigma M} = \frac{1}{2 g'} \left[
\frac{1}{v'_s}(\partial_\tau
{\bf n})^2 + v'_s (\partial_x {\bf n})^2 \right],
\eea
with
\bea
g' = \frac{4}{(K+1) S},~~~~~~v'_s = \frac{4 a_0 J K S}{(K+1)},
\eea
and the topological term
\bea
{\cal L}_{\rm top} = i \frac{K+1}{4}S {\bf n} \cdot (\partial_\tau{\bf n}
\times \partial_x{\bf n}).
\eea

As in the uniform 1D Heisenberg antiferromagnet,
the topological term is proportional to the
Pontryagin density of the hedgehog fields in two dimensions and,
similarly to the antiferromagnet, it is straightforward to show that
for $d > 1$ the topological term is identically zero.
The equations of motion are readily obtained from
(\ref{ferrimagnetaction})
which gives
\bea
M_0 \left({\bf n} \times \partial_\tau {\bf n}\right)_\alpha
&+& \frac{1}{g' v'_s} \left(-\partial^2_\tau +
{\bf n}\cdot\partial^2_\tau{\bf n}\right) {\bf n}_\alpha \nonumber\\&&
+ \frac{v'_s}{g'} \left(-\nabla^2 + {\bf n}\cdot\nabla^2{\bf n}\right)
{\bf n}_\alpha = 0,
\label{eqnsofmotion}
\eea
where a Lagrange multiplier has been used to enforce
the constraint ${\bf n}({\bf r})\cdot{\bf n}({\bf r}) = 1$.
Using the linearized approximation, it is then straightforward to obtain
the spin-wave dispersion relations
\bea
\hbar \omega^{\pm}_k = \frac{4 J K |1-K| S}{(1 + K)^2}
\left[ \sqrt{1 + a_0^2 \frac{(1+K)^2}{(1-K)^2} k^2} \pm 1 \right],
\eea
which agrees well with the dispersion relations obtained from spin-wave
theory.

An interesting new feature of the NL$\sigma$M-type formulation of the
ferrimagnetic chain is a topological term with $\theta = (K + 1) \pi S$,
which gives the familiar result $\theta = 2 \pi S$ when $K = 1$. As can be
easily seen, $\theta$ is
always equal to one of $0$, $\pi$, or $\pm \pi/2$ (mod $2 \pi$) for
ferrimagnetic Heisenberg chains. A
question that immediately comes to mind is whether (in analogy with the
Heisenberg antiferromagnet) the topological term for the
ferrimagnetic chain leads to markedly different behavior for different
values
of $\theta$ or not. The answer is that it does not, and this is due to the
fact that the ultimate long-wavelength behavior of the ferrimagnetic chain
is governed by the ferromagnetic term in (\ref{ferrimagnetaction}) and the
presence of the topological term leads to no surprises in this case. This
is
consistent with fact that
the zero-temperature ground state of the Heisenberg ferrimagnet is ordered
and does indeed correspond to a state with staggered spins and constant
long-range
spin-spin correlations \cite{pati}.
One may show this explicitly by making use of a renormalization group
approach for action (\ref{action3.2}), i.e., the ferrimagnets have a
stable ferromagnetic fixed point.
However, (periodically) alternating   
quantum spin chains other than the Heisenberg ferrimagnet
may possess a topological term which can dramatically affect the
long-wavelength behavior
of the spin chain due to (strong) quantum fluctuations at zero
temperature.
An example which illustrates this type of behavior is the model studied in
\cite{fukui} which consists of a periodic array of two kinds of spins.
Models such as those studied in \cite{fukui}, where the topological
$\theta$ term does indeed play a role,  are likely to lead to novel 
interesting behavior in quantum magnetic systems with useful
practical applications.

\section{Quantum Phases}
\label{QPT}

The semiclassical methods that we have used in the previous sections 
to explore the physics of quantum ferrimagnets are, generally 
speaking, effective descriptions which are based on approximations 
in the continuum limit. However, although these powerful techniques 
have enabled us to learn much, to understand the existence and 
cross-over of (quantum) phases in quantum ferrimagnets other 
approximations need to be made. In this section we will 
present a (continuum) formulation based on dimerization which is 
paricularly powerful and well-suited for a description of quantum 
phases.   

The energy eigenstates of each dimer are specified through the total
angular momentum, $\ell=|S_1 - S_2|, |S_1 - S_2|+1, \dots, S_1 + S_2$, and 
the ground state and the first excited state of a dimer
separated by a gap proportional to the exchange coupling $J$.
In contrast to a dimer with identical spins, whose ground
state is unique, the ground state of a dimer with alternate spins is
degenerate. This degeneracy of the ground state dramatically alters
the low energy physics of the ferrimagnets; making them behave more like
ferromagnets rather than antiferromagnets.
However, turning on the interaction between the dimers removes the
degeneracy 
(allowing the localized states to hop to neighboring dimers) and a gapless
band for the lowest $\ell$ appears.
This corresponds to the gapless spin wave mode 
(as obtained in the preceeding sections) for the intra-band transitions
with the dispersion law $\omega_k \sim k^2$.
On the other hand, for isolated (non-interacting) dimers the gap between
the lowest
angular momentum ($\ell=|S_1 - S_2|$) and the first
excited state ($\ell+1$) is robust and persists even for an alternate spin
chain with a
uniform interaction between the dimers.
This corresponds to the gapped mode of the spin waves with inter-band
transitions. This gap, which is
proportional to $J|S_1 - S_2|$, results from transitions between
different bands with different angular momenta $\ell$.
We will return to this point in the next section (Sec.~\ref{QRG}) where 
we show that the ground state of one-dimensional
quantum ferrimagnets is always gapless and ferromagnetic,
no matter what the values of $S_1$ and $S_2$ are, as long as 
$S_1 \neq S_2$ ($K\neq 1$).

The phases of the system at non-zero temperatures can be understood as
follows:
At low temperature the phase of the system can be characterized by gapless
transitions between the states inside the first band and the system
behaves 
like a ferromagnet.
At high temperature the phase of the system is characterized by gapped
transitions from the lower to the higher bands ($\ell \rightarrow \ell+1$) and  
the system exhibits antiferromagnetic features in this phase 
(see Fig.~\ref{Fig1}).
(Note that due to the Mermin-Wagner theorem there can be no long-range
order in ferrimagnetic 
spin chains in either of these cases when $d=1,2$, although at $d=2$
the system exhibits quasi-long range order.)

As we have seen before, the ferrimagnetic spin-waves 
consist of both gapless and gapped modes. In the 
low temperature limit the latter mode is irrelevant. At $d=1$ and $T=0$
the low (high) energy physics of quantum ferrimagnets is effectively 
like that of a ferromagnet (antiferromagnet) which is formed by the chain
of (dimerized) unit cells with magnetic moment $M_0(=|S_1 - S_2|)$.
This will be shown explicitly by a renormalization group calculation in 
the following section.
Applying an external magnetic field, $h$, 
leads to gapped spin-waves; in which case 
the energy cost for ferromagnetic transitions is proportional to the
Zeeman splitting factor.
In fact, the dispersion relation for spin waves based on the classical
ground state of the ferrimagnets (classical N\'eel state) 
can be obtained from Eq.~(\ref{R8})
\begin{eqnarray}
\hbar\omega^{\pm}(k) &=& \pm \left(\frac{-hS}{2} ~ + ~
\frac{|K-1|}{2} JSz \right) \nonumber\\&& +
\frac{1}{2}JSz
\sqrt{(K - 1)^2 + 4 K[1-\frac{1}{z^2} |\gamma(k)|^2]}.
\label{sph}
\end{eqnarray}
As one can see, the effect of the external magnetic field is to
suppress the ferrimagnet gap. Therefore the ground state of 
the ferrimagnet corresponds to the {\em staggered}  configuration of spins on 
sublattices $A$ and $B$, unless $h_{c1} \geq 2Jz |K-1|$. 
At this point the staggered
state becomes unstable against the {\em canted} phase (spin-flop state) of 
the spins when the spectrum becomes soft at $k=0$. When the external magnetic
field exceeds $h_{c2}$, the system will be in a saturated
ferromagnetic phase, with a quantized magnetization per unit cell.
This is shown in Fig. \ref{newfig2}.  
The $h_{c2} \equiv 2Jz(1+K)$ is 
obtained by using the dispersion relation of the spin waves based on the fully 
polarized state of the ferrimagnets. 
It is the lower-bound of the external magnetic
field, in the sense that the spin waves become soft at $k=0$.
Clearly, this reveals  
similarities between the ferrimagnets and the $XXZ+h$ models \cite{assa}.  
However, we will not pursue this interesting connection any further and leave 
it for future investigations.

As has been shown here, the problem of a $d$-dimensional quantum
ferrimagnet can be mapped onto a classical $d+1$-dimensional non-linear
$\sigma$-model with or without a topological term for spin systems
in $d=1$ or $d\neq 1$ respectively.
Turning back to Eq.(\ref{ferrimagnetaction}), let us now reparameterize
the (3-component) unit verctor ${\bf n}$ as:
\begin{equation}
{\bf n}({\bf r}) = \left(\sin\theta({\bf r})\cos\varphi({\bf r}),
\sin\theta({\bf r})\sin\varphi({\bf r}), \cos\theta({\bf r})\right),
\end{equation}
where ${\bf r}$ denotes the position of ${\bf n}$ in 
$d+1$ dimensions.
On the other hand, 
as we have seen previously through the spin wave calculations,
when $h_{c1} < h < h_{c2}$ the ferrimagnet is in the canted phase,
which contains an in-plane gapless
mode---even in the presence of an external perpendicular magnetic field---
and this gapless mode can significantly affect the low energy physics 
\cite{Subir,Cote,skyrmelattice}. In this connection, it is 
convenient to think of $\theta({\bf r})$ as a
`gap field variable' when the $O(3)$ symmetry is broken.
We can then integrate out the gapped mode and obtain  
an effective Lagrangian for the in-plane gapless mode (on a
$d$-dimensional lattice):
\begin{eqnarray}
{\cal L}_{\rm eff} &=& \sum_j \left\{ M_0 \partial_t \varphi_j +
\frac{1}{2g'v_s'} (\partial_t \varphi_j)^2 \right\} \nonumber\\&&
+ \frac{2v_s'}{g'} \sum_{<i,j>} \cos(\varphi_i - \varphi_j).
\label{effl}
\end{eqnarray}
The first term in (\ref{effl}) is the usual ferromagnetic 
Berry's phase and the last term corresponds to the classical XY
model.
In the continuum limit, the XY model is simply the Laplacian of
the field $\varphi({\bf r})$ and by introducing momenta 
conjugate to $\varphi_j$
\begin{equation}
M_j \equiv \frac{\partial{\cal L}}{\partial(\partial_t\varphi_j)} =
M_0 + \frac{1}{g'v_s'} \partial_t\varphi_j,
\end{equation}
one obtains the effective Hamiltonian
\begin{eqnarray}
{\cal H}_{\rm eff} &=& \frac{g'v_s'}{2} \sum_j \left(\frac{1}{i}
\frac{\partial}{\partial\varphi_j} - M_0\right)^2 \nonumber\\&&
- \frac{2v_s'}{g'} \sum_{<i,j>} \cos(\varphi_i - \varphi_j),
\label{effl1}
\end{eqnarray}
where we have used the quantum mechanical representation of the number
of the flipped magnetic moments per unit cell
\begin{equation}
|M_j\rangle \propto \int d\varphi_j e^{i\varphi_j M_j} |\varphi_j\rangle,
\end{equation}
which is canonically conjugate
to the classical in-plane orientation \cite{Cote}.
The Hamiltonian (\ref{effl1}) is the boson Hubbard model (the quantum
rotor with a non-zero minimum angular momentum).
The superfluid (Mott-insulating) phase of bosons can be obtained
when $g' \ll 2~(g' \gg 2)$.
When $g' \rightarrow \infty$, and
for any value of $M_0$, except odd integr $2M_0$'s,
the ground state consists of the gapped Mott-insulating phase,
and the classical ordered phase is destroyed by quantum fluctuations.
It then follows that at $T=0$, a ferrimagnet with
arbitrary (but unequal) spins $S_1$ and $S_2$ is
actually in a superfluid phase of the condensated bosons. 
Within this description we can also see that 
at $d=2$ quantum ferrimagnets exhibit  
critical behavior where a Berezinskii-Kosterlitz-Thouless
phase transition \cite{KT} takes place at $T_{BKT} = \pi v'_s/g'$.
Furthermore, the magnetization $M_0 (\equiv \langle S_1 \rangle + 
\langle S_2 \rangle)$ does not have a quantized value and varies
continuously as a function of the couplings $g'$ and the 
external magnetic field, $h$. This is the partially magnetized
phase and/or the spin-flop phase.

\section{The Quantum Renormalization Group Approach}
\label{QRG}

As has been discused in the previous section, it is possible to capture
some important features 
of the physics of quantum ferrimagnets through effective (field) theories
based on dimerization 
within the semiclassical approximation. In this section we wish to show
that an effective theory 
based on dimerization and not necessarily reliant on some underlying
assumptions (for example the 
large-$S$ limit) in semiclassical treatments can also reveal much about
the behavior of these systems. 
In particular, we will show that a quantum renormalization group (QRG)
study of ferrimagnetic 
quantum spin chains yields good quantitative and qualitative results. As
opposed to other 
powerful techniques,  
such as the density matrix renormalization group (DMRG) \cite{white}, the
QRG approach is 
much less complicated and intuitively straightforward. The QRG is based on
the much familiar 
standard block renormalization and is also analytically quite tractable
\cite{pfeuty} (see also 
\cite{miguel} for a modern treatment and \cite{abdollah} for some recent
applications). 

To apply the QRG technique one begins by dividing up the spin lattice into
small blocks (see Fig. \ref{Fig2}) 
and obtains the lowest-energy states ($\ket{\al}$) of each isolated
block for a particular
boundary condition---for example, for the dimerized states that we will
use here open boundary 
conditions for isolated blocks are chosen. The effect of inter-block
interactions is then taken into account
by constructing an effective Hamiltonian $H^{eff}$
which now acts on a smaller
Hilbert space ${\cal H }^{eff}$ embedded in the original one. 
In this new Hilbert space, each 
of the former blocks is treated as a single site and  
the technical way of implementing this idea is to construct an embedding
operator $$ Q: {\cal H}^{eff} \longrightarrow {\cal H } $$  and 
a truncation operator 
$$ Q^{\dagger}: {\cal H} \longrightarrow {\cal H }^{eff} $$
by demanding the commutativity of the following diagram\cite{miguel}:
$$
\begin{array}{c}
{\cal H}^{eff} \stackrel{Q}{\longrightarrow}  {\cal H } \\
H^{eff}  \downarrow  \hspace{1.5cm} \downarrow \hspace{0.5cm} H \\
{\cal H}^{eff} \stackrel{Q}{\longrightarrow}  {\cal H } \\
\end{array}
$$
{\it i.e.} such that: $Q H^{eff} = H Q$.
>From the last relation one obtains the effective Hamiltonian as :
\beq
\label{rg1}
H ^{eff} = Q^{\dagger} H Q.
\eeq
Note that the operators $Q$  and $Q^{\dagger}$ satisfy the 
relation $ Q^{\dagger} Q = 1_{{\cal H}^{eff}} $ 
but $ Q Q^{\dagger } \ne 1_{{\cal H }} $. 
More precisely, one divides $H$ as $H=H^{B}+H^{BB}$ (see Fig. \ref{Fig2})  
where $H^{B}=\sum_{I}h_{I}^{B}$ is the sum of block Hamiltonians
and $H^{BB}=\sum_{<I,J>}h_{I,J}^{BB}$ is the sum of inter-block
Hamiltonians.
Then $Q=\prod_{I}Q_{I}$ with $Q_{I}$ given by
\beq
\label{rg2}
Q_{I}=\sum_{\al=1}^{m}\ket{\al}\bra{\al},
\eeq
where $m$ is the number of low energy states that are kept.
Note that since each $Q_{I}$ acts trivially on all the other blocks $J \ne
I$, $[Q_{I},Q_{J}]=0$.

Let us now specialize this construction to a one-dimensional quantum
ferrimagnet with $S_1=1/2$ and $S_2=1$.
Here, each cell labeled by $i$ is considered as an isolated block in the
renormalization group 
procedure. The block Hamiltonian ($H^B$) and the inter-block
Hamiltonan ($H^{BB}$) can be written as
\beq
\label{rg3}
H^B=J\sum_{i=1}^{N} {\bf S}_{1,i} \cdot {\bf S}_{2,i},
\eeq
\beq
\label{rg4}
H^{BB}=J \delta \sum_{i=1}^{N} {\bf S}_{2,i} \cdot {\bf S}_{1,i+1},
\eeq
where we also assume that the spin chain obeys the periodic boundary
conditions: 
${\bf S}_{1(2),N+1}={\bf S}_{1(2),1}$.
The Hilbert space of each isolated block consists of a $S=3/2$ and
$S=1/2$  multiplet corresponding to $(J/2)$ and $(-J)$ in energy.
We choose the $S=1/2$ multiplet base kets as the kept states in the
QRG procedure to find the effective low-energy Hamiltonian.
(The higher energy states of the $S=3/2$ multiplet are neglected.)
If we specify these states by $\ket{S,M}$, where $S$ is the total spin
and $M$ is its z-component, we can construct the embedding operator
of each block as
\beq
\label{rg5}
Q_i=\ket{\frac{1}{2},+\frac{1}{2}}\bra{+}
\;+\;\ket{\frac{1}{2},-\frac{1}{2}}\bra{-},
\eeq
where $\ket{+}$ and $\ket{-}$ are the renamed base kets in the effective
Hilbert space of each block.

By using Eq.(\ref{rg1}) one can find the effective spin operators of the
renormalized chain:
\bea
\label{rg6}
Q_i^{\dagger} {\bf S}_{1,i}^{\al} Q_i = \frac{-1}{3} {\bf S'}_i^{\al},
\nonumber \\
Q_i^{\dagger} {\bf S}_{2,i}^{\al} Q_i = \frac{4}{3} {\bf S'}_i^{\al},
\eea
where $\al$ stands for ${+,-,z}$ and ${\bf S'}$ is the matrix
representation
of the spin 1/2 operator in the effective Hilbert space of the
block-renormalized chain.
The effective low-energy Hamiltonian ($H_{low}^{eff}$)
can be obtained from 
\beq
\label{rg7}
H_{low}^{eff}=Q^{\dagger} H_{2N}(J,\delta) Q=Q^{\dagger} H^{B} Q
+Q^{\dagger} H^{BB} Q.
\eeq

Using ${\bf S}_{2,i} \cdot {\bf S}_{1,i+1}=
\frac{1}{2}({\bf S^+}_{2,i}{\bf S^-}_{1,i+1}+{\bf S^-}_{2,i}{\bf
S^+}_{1,i+1})
+{\bf S^z}_{2,i}{\bf S^z}_{1,i+1}$ along with Eq.(\ref{rg6}) one has
\beq
\label{rg9}
Q_{i+1}^{\dagger}Q_i^{\dagger}{\bf S}_{2,i} \cdot {\bf S}_{1,i+1} Q_i
Q_{i+1}
=\frac{-4}{9} {\bf S'}_{i} \cdot {\bf S'}_{i+1}, 
\eeq
from which it follows that
\beq
\label{rg10}
Q^{\dagger} H^{BB} Q=\frac{-4}{9} J \delta
\sum_{i=1}^N {\bf S'}_i \cdot {\bf S'}_{i+1}.
\eeq
>From this last equation and the equation 
\beq
Q^{\dagger} H^B Q = -N J,
\label{rg10.1}
\eeq
we see that the effective low-energy Hamiltonian $(H_{low}^{eff}(1/2,1))$
describing a ($S_1=1/2,S_2=1$) ferrimagnetic Heisenberg chain
is the same as that of a spin 1/2 Heisenberg ferromagnet with a
renormalized
coupling constant, $J_{low}^{(1/2,1)}=(4/9) J\delta > 0$, as given by
\beq
\label{rg11}
H_{low}^{eff}(1/2,1)=-NJ - J_{low}^{(1/2,1)} \sum_{i=1}^N {\bf S'}_i \cdot
{\bf S'}_{i+1}
\eeq
This result is in complete agreement with the results of the previous
sections, confirming that the effective low-energy behavior of a
ferrimagnetic chain corresponds to a gapless ferromagnet. The approximate
ground state energy per cell which is obtained from Eq.~(\ref{rg11}) is
$(1/N)E_0=-J(1+(\delta/9))$.
As a result of this effective description, one can also obtain the 
ferrimagnetic gapless mode dispersion relation ($\omega \sim k^2$) 
by using the
dispersion relation of the $S=1/2$ Heisenberg ferromagnet in
Eq.~(\ref{rg11}).

In addition to considerations that have to do with the ferrimagnetic
ground state, we 
can also gain some insight about the low-energy excitations of the quantum
ferrimagnets  
by constructing another
embedding operator ($Q$) which gives the effective Hamiltonian for the
excited spectrum.
The first excited state is obtained by considering one dimer in the
$S=3/2$ multiplet Hilbert space. In this case the embedding operator of
the excited dimer is constructed by the base kets of the $S=3/2$ multiplet
and the
other ones remain as before. The resulting Hamiltonian obtained from the
QRG 
process in this case is given by
\bea
\label{rg12}
H_{ex}^{(1)}&=&(-NJ+\frac{3}{2}J)+
\frac{2}{9} J \delta ({\bf S'}_{j-1} \cdot {\bf S''}_{j}+{\bf S''}_{j}
\cdot {\bf S'}_{j+1}) \nonumber \\
&-&\frac{4}{9}J \delta \sum_{i\neq j-1,j} {\bf S'}_{i} \cdot {\bf
S'}_{i+1},
\eea
where $S'=1/2$ and $S''=3/2$.
The first term in (\ref{rg12}) contributes to the gap and the second one
is
responsible for antiferromagnetic interactions.
(Therefore the gapped excitations exhibit antiferromagnetic interactions).
If we continue this procedure by considering more dimers in the $S=3/2$
multiplet Hilbert space, other excited states are obtained and
finally for the highest energy spectrum we find
\beq
\label{rg13}
H_{ex}^{(N)}=\frac{N}{2} J + \frac{2}{9} J \delta \sum_{i=1}^N
{\bf S''}_{i} \cdot {\bf S''}_{i+1}.
\eeq
Therefore for sufficiently high energies the physics of the
($S_1=1/2,S_2=1$)
ferrimagnetic Heisenberg chain corresponds to a spin $3/2$ Heisenberg
antiferromagnet which is separated by a finite gap from the gapless
ferromagnetic branch.

To further substantiate our results we have also calcualted
the (average) magnetization per spin in the unit cells and the spin-spin
correlation
functions. In the QRG formalism the ground state $\ket{0}$ is replaced
by $Q\ket{0'}$, where $\ket{0'}$ is the ground state of the effective
Hamiltonian Hilbert space. Therefore the magnetization, $m_l$, per site
for
each spin is 
\bea
\label{rf1}
m_l&=&\bra{0}\frac{1}{N} \sum_{i=1}^{N} {\bf S}_{l,i}^z \ket{0}, \nonumber
\\
&=&\frac{1}{N}\sum_{i=1}^N \bra{0'}Q_i^{\dagger} {\bf S}_{l,i}^z Q_i
\ket{0'},
\eea
with $l = 1, 2$ for $S_1$ and $S_2$ respectively. (Note that the spin-wave
sum rule,
$M_0 \equiv m_1 + m_2 = |S_1 - S_2|$, is satisfied by Eq.~(\ref{rf1}).)
We have summerized the results of calculations of magnetization, energy
gap($\Delta$), and
the ground state energy which has been obtained by using the QRG
techniques (this section),
spin-wave analysis (Sec.~II and III, \cite{pati}), DMRG (S.~K.~Pati {\it
et al.} \cite{pati}), 
and quantum Monte-Carlo (QMC) calculations (S.~Brehmer, {\it et al.}
\cite{pati}) in Table.~1
for a quantitative comparison. As can be seen from Table.~1, the QRG
results are closer
to those found from the spin-wave
approach which is, however, not as accurate as the DMRG and QMC results.
It is known
\cite{abdollah} that in the QRG method the number of kept states and the
lenght
of blocks is responsible for the accuracy of the results due 
to the boundary effects in 
isolated blocks. However, this loss of accuracy, as compared to DMRG and
QMC, {\it e.g.},
is compensated for by straightforward analytically tractable results
which explicitly exhibit renormalization group flows towards the fixed
points.
In the example discussed here we were able to obtain the real-space
representation 
of the effective low-energy Hamiltonian explicitly to study the physical
properties of 
ferrimagnetic Heisenberg chains.

The results of our calculation of the correlation function 
between the z-components of the $S=1/2$ spins and an arbitrary spin on the 
chain is plotted in Fig.~\ref{Fig3}.
We have also plotted the QMC results (S.~ Brehmer, et.al \cite{pati}) for
comparison. The QRG results are in complete qualitative agreement with the
QMC
data, but there are slight quantitative differences between the two
approaches. 
In particular, in the long-wavelength limit the QRG approach agrees more
closely 
with the spin-wave theory than QMC calculations.  
However, both QRG and QMC yield a correlation lenght which is smaller 
than a lattice spacing.

In addition to supporting the ideas that were discussed in the previous
sections, 
the results in Table.~1 and Fig.~\ref{Fig3} confirm that the QRG method
leads to 
good qualitative and quantitative results while maintaining an overall 
anlytical simplicity. In fact, besides the ($1,1/2$) that we just
discussed in detail, 
we have also studied two other cases ($S_1=1/2,S_2=3/2$) and
($S_1=1,S_2=2$) with comparable ease. 
We find that the same general picture holds for these two
cases as well. Then the effective low-energy Hamiltonian for the 
($S_1=1/2,S_2=3/2$) is given by
\beq
\label{rg14}
H_{low}^{eff}(1/2,3/2)=-(\frac{5}{4})NJ
- J_{low}^{(1/2,3/2)} \sum_{i=1}^N {\bf S}_{i}^{(1)} \cdot {\bf
S}_{i+1}^{(1)},
\eeq
where $J_{low}^{(1/2,3/2)}=(5/16) J\delta > 0$ and ${\bf S}^{(1)}$ is a
spin 1.
For the ($S_1=1,S_2=2$) case we find 
\beq
\label{rg15}
H_{low}^{eff}(1,2)=-3NJ
- J_{low}^{(1,2)} \sum_{i=1}^N {\bf S}_{i}^{(1)} \cdot {\bf S}_{i+1}^{(1)}
\eeq
where $J_{low}^{(1,2)}=(3/4)J\delta > 0$.
One can use (\ref{rg14}) and (\ref{rg15}) to obtain the ferromagnetic
gapless dispersion relations ($\omega \sim k^2$) for these ferrimagnetics. 
For example, the approximate
ground state energy and the magnetization per spin are
\bea
\label{rg16}
\frac{1}{N} E_0(1/2,3/2) =\frac{-5}{4} J (1+\frac{\delta}{4}), \nonumber
\\
m(\frac{1}{2})=-0.25\;\;;\;\;m(\frac{3}{2})=1.25,
\eea
and
\bea
\label{rg17}
\frac{1}{N} E_0(1,2) =-3 J (1+\frac{\delta}{4}), \nonumber \\
m(1)=-0.5\;\;;\;\;m(2)=1.5.
\eea

The general conclusion that emerges from the QRG calculations that we have
performed 
is that the effective low-energy Hamiltonian of a ($S_1,S_2$)
ferrimagnetic 
Heisenberg chain is the same as that of a spin $S=|S_1-S_2|$ ferromagnetic
chain
with a gapless mode. Our QRG study also exhibits the cross-over to
antiferromagnetic 
behavior at a finite energy gap which was also obtained in the previous
sections 
by using semiclassical approximations. A more refined analysis of 
quantum ferrimagnetic spin system within 
the QRG framework and at finite temperature can, in principle, lead to
quantitative 
predictions in the region where the ferromagnetic-antiferromagnetic
cross-over takes 
place. In view of the complexity of the problem through using the
continuous renormalization 
group approach---even when combined with semiclassical approximations such
as the large-$S$ 
limit, {\it e.g.}---it would be very instructive to examine these
questions in future investigations  by using QRG 
techniques.
  
\section{Conclusion}
In this paper we have presented an extensive study of quantum ferrimagnets 
in one, two, and three dimensions at both zero and finite temperature 
by using a variety of techniques. These include the 
spin-wave theory semiclassical approximation, the continuum non-linear 
$\sigma$-model type effective field theory, a formulation based on the
quantum rotor models, and the real space quantum renormalization group
approach. We find a rich phase diagram for quantum ferrimagnets with
nearest-neighbor interactions at $T=0$ and $T>0$. The effective
description of these systems considerably simplifies in the long- and
short-wavelength limits. In the long-wavelength limit quantum ferrimagnets
behave as ferromagnets with a (quantized) magnetization $|S_1 - S_2|$ per
unit cell, whereas in the short-wavelength limit they exhibit strong
antiferromagnetic features.
We also demonstrate the existence of an intermediate continuous phase
transition
to a partially polarized magnetized state in quantum ferrimagnets.
Similarly to the usaul ferromagnets and antiferromagnets, one-dimensional
quantum ferrimagnets do not exhibit long range order at non-zero
temperatures; however, two-dimensional ferrimagnets 
show a quasi-long range order at non-zero temperature and a 
Berezinskii-Kosterlitz-Thouless type of phase transition can be seen.
Three-dimensional quantum ferrimagnets behave differently from their one-
and two-dimensional counterparts and exhibit both ferromagnetic and
antiferromagnetic long range order, the relative strength of which depends
on the temperature and the ratio $S_1/S_2$ of neighboring spins.
At $T=0$ quantum ferrimagnets show long range order at any
dimension. When the directions of $S_1$ and $S_2$ are opposite, 
both the magnetization $M_0 = |S_1 - S_2|$ and
the staggered magnetization $M^{Neel}$ are finite and can be
considered 
as order parameters. 
And finally, We have shown that
the effective low-energy Hamiltonian of a ($S_1,S_2$)
ferrimagnetic 
Heisenberg chain is the same as that of a spin $S=|S_1-S_2|$ ferromagnetic
chain.
The methods and applications discussed in this paper
have enabled us to obtain good quantitative and qualitative results and
can be extended to study more general quantum ferrimagnets. 

\section{Acknowledgement}
It is a pleasure to thank Assa Auerbach, Steve Girvin, Hans Hansson,
Anders Karlhede, and Subir Sachdev for comments, correspondence, and
discussions. 
H.H. is suported by the US Department of Energy 
under contract DE-FG02-91ER40676. 


\begin{table}
\caption{A comparison of the magnetization per site ($m_l$),
the energy gap ($\Delta$), and the 
ground state energy per unit cell ($E_0/J N$) as obtained from QRG,
spin-wave (SW) theory, DMRG, and QMC
calculations.}
$$
\begin{array}{|c|c|c|c|c|}\hline
& QRG & SW & DMRG & QMC \\ \hline
m(S=\frac{1}{2}) & -0.166 & -0.195 & -0.292 & -0.290 \\ \hline
m(S=1) & 0.666 & 0.695 & 0.792 & 0.790 \\ \hline
\Delta/J & 1 & 1 & 1.279 & 1.767 \\ \hline
E_0/J N & -1.111 & -1.436 & -1.454 & -1.455 \\ \hline
\end{array}
$$
\label{table1}
\end{table}

\begin{figure}
\caption{
The finite temperature phase diagram for the 3-dimensional
quantum ferrimagnets.
}
\label{Fig1}
\end{figure}

\begin{figure}
\caption{
The magnetization per unit cell ($M_0$) as a function of the
external magnetic field. When $h_{c1} < h < h_{c2}$ the ferrimagnet is 
partially polarized. At $h = h_{c2}$ the external field leads to a saturated
polarized (ferromagnetic) state.}
\label{newfig2}
\end{figure}

\begin{figure}
\caption{
Block decomposition of the ferrimagnetic chain into unit cells.
}
\label{Fig2}
\end{figure}

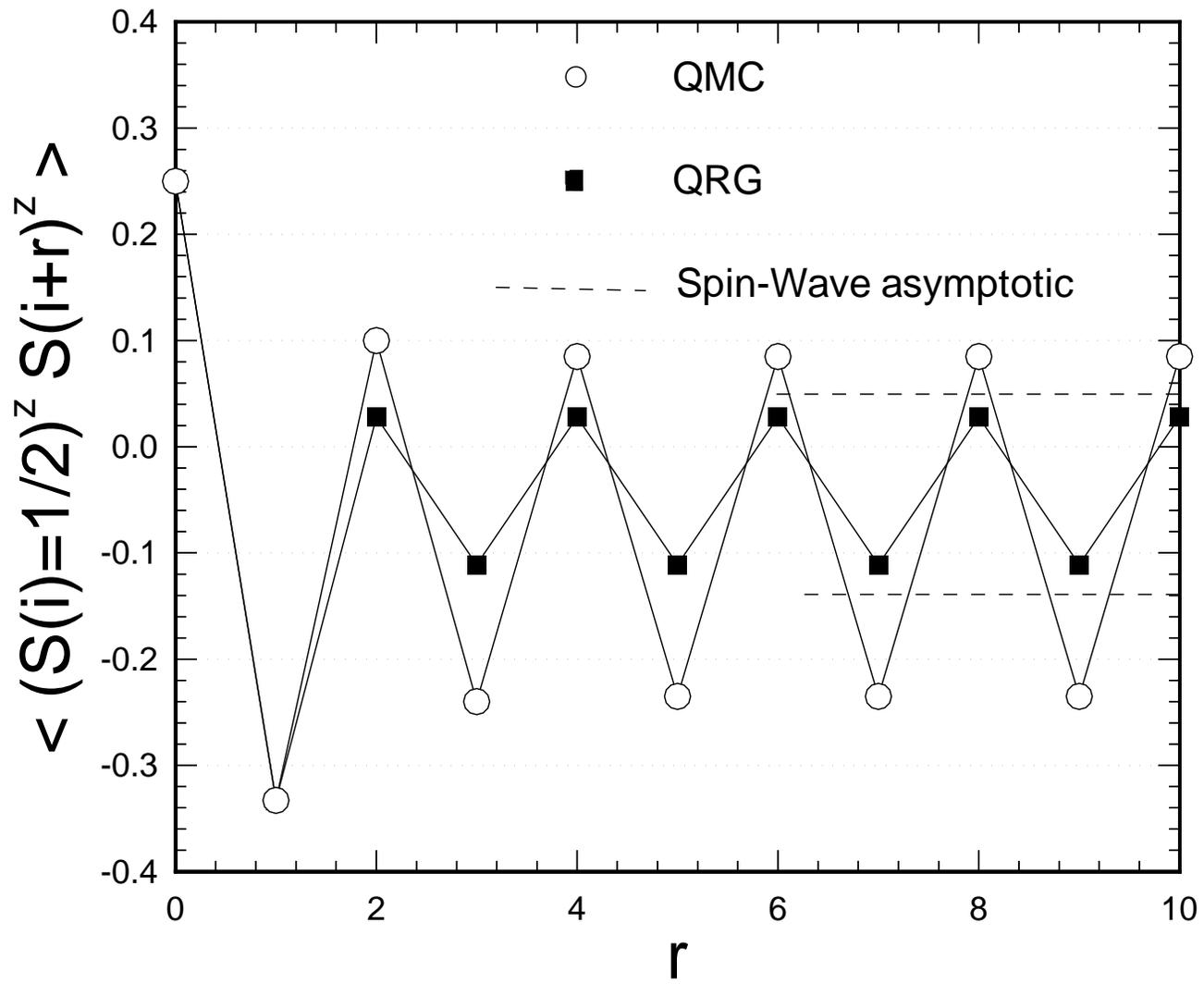
\begin{figure}
\caption{
The $\langle S_{1z} S_{2z} \rangle$ correlation function vs. distance 
as obtained from QRG and QMC. Here $S_{1z} = 1/2$ and $S_{2z}$ is the 
$z$-component of spin on an arbitrary site.
}
\label{Fig3}
\end{figure}

\end{document}